\documentclass[usenatbib]{mn2e}
\usepackage{graphicx}

\newcommand{\kpch}{{h^{-1}\rm kpc}}

\title[Assessing the JAM method with the Illustris simulation]{Assessing the Jeans Anisotropic Multi-Gaussian Expansion method with the Illustris simulation}
\author[Hongyu~Li et al.]{Hongyu~Li$^{1,2}$\thanks{E-mail: hyli@nao.cas.cn}, Ran~Li$^{1, 8}$\thanks{E-mail: ranl@bao.ac.cn}, Shude~Mao$^{3,1,4}$, Dandan~Xu$^{5}$, R.~J.~Long$^{1,4}$ \and and Eric Emsellem$^{6,7}$\\
$^{1}$National Astronomical Observatories, Chinese Academy of Sciences, 20A Datun Road, Chaoyang District, Beijing 100012, China\\
$^{2}$University of Chinese Academy of Sciences, Beijing 100049, China\\
$^{3}$Physics Department and Tsinghua Centre for Astrophysics, Tsinghua University, Beijing 100084, China\\ 
$^{4}$Jodrell Bank Centre for Astrophysics, School of Physics and Astronomy, The University of Manchester, Oxford Road, Manchester M13 9PL, UK\\
$^{5}$Heidelberg Institute for Theoretical Studies, Schloss-Wolfsbrunnenweg 35, 69118 Heidelberg, Germany\\ 
$^{6}$Universit\'e Lyon 1, Observatoire de Lyon, Centre de Recherche
Astrophysique de Lyon and Ecole Normale Sup\'erieure de Lyon, 9 avenue\\ 
Charles Andr\'e, F-69230 Saint-Genis Laval, France\\
$^{7}$European Southern Observatory, Karl-Schwarzschild-Str. 2, 85748 Garching, Germany\\
$^{8}$Key laboratory for Computational Astrophysics, National Astronomical Observatories, Chinese Academy of Sciences, Beijing, 100012, China}
\begin{document}

\date{Accepted 2015 October 29. Received 2015 October 28; in original form 2015 September 16}

\pagerange{\pageref{firstpage}--\pageref{lastpage}} \pubyear{2015}

\maketitle
\label{firstpage}
\begin{abstract}
We assess the effectiveness of the Jeans-Anisotropic-MGE (JAM) technique with a state-of-the-art cosmological hydrodynamic simulation, 
the Illustris project.  We perform  JAM modelling on 1413 simulated galaxies with stellar mass $M^* > 10^{10}M_{\odot}$, and construct an axisymmetric dynamical model for each galaxy.  Combined with a Markov Chain Monte Carlo (MCMC) simulation, we recover the projected root-mean-square velocity ($V_{\rm rms}$) field 
of the stellar component, and investigate constraints on the stellar mass-to-light ratio, $M^*/L$, and the fraction of dark
 matter $f_{\rm DM}$ within 2.5 effective radii ($R_e$).  We find that the enclosed total mass within $2.5 \: \rm R_e$ is well constrained to
  within $10\%$. However, there is a degeneracy between the dark matter and stellar components with correspondingly larger individual errors. The 1$\sigma$ scatter in the recovered $M^*/L$ is  $30-40\%$ of the true value. The accuracy of the recovery of $M^*/L$ depends on the triaxial shape of a galaxy. There is no significant bias for oblate galaxies, while for prolate galaxies the JAM-recovered stellar mass is on average 18\% higher than the input values. We also find that higher image resolutions alleviate the dark matter and stellar mass degeneracy and yield systematically better parameter recovery.
\end{abstract}

\begin{keywords}
galaxies: kinematics and dynamics - galaxies: formation - 
galaxies: evolution -  dark matter - galaxies: structure
\end{keywords}

\section{Introduction}

With the increasing availability of Integral Field Units (IFUs), more and more nearby galaxies with IFU data are becoming available, e.g. ATLAS$^{\rm 3D}$ \citep{b10}, CALIFA \citep{Sanchez2012}, SAMI \citep{b12}, MaNGA \citep{b11}. The kinematic information so offered can be used with dynamical modelling techniques to investigate stellar kinematics, galaxy structure, galaxy evolution, merger processes, mass distributions and so on. Existing methods for constructing  dynamical models include distribution function based methods (e.g. \citealt{b14,b15}), orbit methods based on \citet{b16} (e.g. as implemented in \citealt{zhao1996,Hafner2000,Bosch2008,Wang2013}), particle method based on \citet{Syer1996} (e.g. as implemented in\citealt{Lorenzi2007,Long2010,Zhu2014,Hunt2013}), and  moment-based methods (e.g. \citealt{b6}), which find  solutions of the Jeans equations. Clearly there will be existing methods which do not fit neatly in to this categorisation: for example, two particle based methods not based on \citet{Syer1996} are \citet{Yurin2014} and \citet{Rodionov2009}.
 Each method has its own strengths and can be applied under different conditions. For action-angle based distribution functions, it depends on having the ability to find the actions \citep{Sanders2015}. Orbit-based or particle-based methods are more accurate and flexible, but are time consuming and challenging to apply to large samples (e.g. for MaNGA, there will be $\sim$10000 galaxies available by the end of the survey). In order to take advantage of a large sample and obtain statistical results, a moment-based method like the Jeans Anisotropic Multi-Gaussian Expansion (JAM) method \citep{b6} may be a good compromise, at least in terms of balancing computational efficiency against potentially restrictive scientific assumptions. 

The JAM  method has already been applied in many studies. In the SAURON project \citep{Zeeuw2002}, it was used to study, for example, galaxy inclination, mass-to-light ratio \citep{b6}, and escape velocity in early-type galaxies \citep{b19}. In ATLAS$^{\rm 3D}$ \citep{b10},  dark matter was also included to study both the stellar mass-to-light ratio and the dark matter fraction for galaxies \citep{b5}.  The stellar initial mass function (IMF) has also been constrained by the stellar mass-to-light ratio predicted by the JAM method \citep{b20}. Similarly, in \citet{b21} and \citet{b22},  JAM modelling was combined with gravitational lensing to investigate the IMF.

While the JAM technique has been extensively used, its accuracy is not well understood. It may have potential biases and large scatters in the parameters being estimated. A good way to assess JAM is using simulated galaxies with known results. In \citet{b23}, they studied the effect of applying  axisymmetrical models to non-symmetrical systems, and obtained  accuracies for some of their parameters. However, they did not include dark matter in their models, and the mock galaxies they used were not taken from a cosmological simulation with extensive sub-grid physics. In reality, galactic structures are complex, and testing with more realistic scenarios and wide ranges of galaxy properties is needed -- thus, the use of cosmologically simulated galaxies in this paper. A similar test has been performed for the Schwarzschild modelling in \citet{Thomas2007}, and the results have been applied to Coma galaxies \citep{Thomas2007b}. In \citet{Thomas2007}, they use merger remnants as mock galaxies to test the accuracy of the recovered stellar mass-to-light ratio, total mass, and their dependance on the viewing angle and galaxy triaxial shape (see Section~\ref{summary} for comparisons with our results).   

The structure of the paper is as follows. In {Section~\ref{method}}, we give a brief introduction to the simulations and the mock data. In {Section~\ref{jam_model}}, we introduce the JAM method. In {Section~\ref{rst}}, we show our results on the bias and degeneracy in $M^*/L$ and $f_{\rm DM}(2.5\: \rm R_e)$ (the fraction of dark matter enclosed within $2.5 \: \rm R_e$).  We summarise and discuss our results in {Section~\ref{summary}} and {Section~\ref{discussion}}.

\section{Simulations and Mock galaxies}\label{method}
\subsection{The Illustris Simulation}\label{sec:simulation}

The Illustris project \citep{Vogelsberger2014a, Vogelsberger2014b, Genel2014, Nelson2015} comprises a suite of
cosmological hydrodynamic simulations carried out with the moving mesh code AREPO \citep{springel2010}. 
The hydrodynamical simulation evolves the baryon component with using a number of sophisticated sub-models including star formation \citep{springel2003},
gas recycling, chemical enrichment, primordial cooling \citep{Katz1996}, metal-line cooling, supernova feedback,
and supermassive black holes with their associated feedback \citep{Matteo2005, Springel2005, Sijacki2007}. For complete details, the reader is referred to \citet{Vogelsberger2013}.  The Illustris simulations reproduce
various observational results, such as the galaxy luminosity function, star formation rate to mass main sequence, and the
Tully-Fisher relation \citep{Torrey2014}. The galaxy morphology type fractions as a function of stellar mass and environment 
also agree roughly with observations \citep{Vogelsberger2014b, Snyder2015}.

In this work, we use the largest simulation (Illustris-1 L75n1820FP) in the Illustris project which contains $1820^3$ dark matter
particles and approximately $1820^3$ gas cells or stellar particles. The snapshot that we use is at redshift 0. The simulation follows the evolution of the
universe in a box of $106.5\: \rm Mpc$ on a side, from $z=46$ to $z=0$. The softening lengths for the dark matter and baryon
components are $1420$ $\rm pc$ and $710$ $\rm pc$ respectively. The cosmological parameters adopted
in the simulations are  $\Omega_m=0.2726$, $\Omega_L=0.7274$, $\sigma_8=0.809$, $h=0.704$ and $n_{\rm s}=0.963$ \citep{Vogelsberger2014a}.

\subsection{Mock galaxies}\label{sec:mock}
In this section we describe how we select our sample and  extract the properties of the simulated galaxies. As we aim to assess the validity of the JAM method, we create ideal observational data to enable us to concentrate on JAM and ignore non-JAM matters. This means we do not incorporate any observational effects (e.g. stellar population, seeing, dust extinction). We describe first the extraction process, since a failed extraction means the galaxy can not be part of the sample, and then the sample selection itself.

First we construct a galaxy's kinematic map as well as its observed image by projecting the galaxy's stellar particles onto  two-dimensional grid cells with grid cell size equal to 2 $\kpch$. The mean velocity and velocity dispersion are obtained by calculating the stellar-mass-weighted mean and standard deviation of the line-of-sight velocity for stellar particles in each grid cell. The stellar surface mass density in each grid  cell is calculated by  dividing the total mass in the cell by the cell area. We then convert this surface density
 map into a brightness map by setting the $M^*/L \equiv 1$ for all cells. It is worth noting that such a unity $M^*/L$ is the reference value of the stellar mass-to-light ratio parameter in the JAM models of our mock galaxies (see Section~\ref{jeans_model}). The surface brightness maps are used to provide the light distribution which is used to derive the stellar mass density as well in solving the Jeans equations. Therefore the resolution of these maps can have a crucial effect on the accuracy of the JAM method. In order to test the effect of surface brightness image resolution, we also use another grid with grid cell size equal to 0.5 $\kpch$ for calculating the brightness map. We refer to these maps as the high-resolution. The grid cell size for kinematic data is always 2 $\kpch$ regardless of the different brightness image resolutions.
 
The observables (surface brightness map, kinematic map) derived above are used as inputs to a galaxy's JAM modelling in order to put constraints on other galaxy properties. To assess the accuracy of JAM recovered  parameters, we need to calculate the true parameter values of the properties of our mock galaxies. The radial distribution of dark matter, stellar and total mass and that of the dark matter fraction are all derived by averaging particle distributions within shells assuming spherical symmetry of the galaxy. Then we fit the dark matter density profile using a generalized NFW model (see Section~\ref{mass_model}) to obtain the true dark matter parameters. 

The true shape and inclination angle for each mock galaxy are derived with the reduced inertia tensor method as implemented  by \citet{Allgood2006}. The tensor is defined as
\begin{equation}
I_{i,j}=\sum_{k\in \mathcal{V}} \frac{x_i^{(k)}x_j^{(k)} } {r_k^2}
\end{equation}
where $r_k$ is the distance measure from the centre of mass to the k-th particle, $x_i^{(k)}$ is the i-th coordinate of the k-th particle and $\mathcal{V}$ is the set of particles of interest.  Assuming that a 
galaxy can be  represented by ellipsoids of axis lengths $a\geq b\ge c$, the axis ratios $q = b/a$ and $s =c/a$ are 
the ratios of the square-roots of the eigenvalues of $I$, and the directions of the principal axes are given by the corresponding 
eigenvectors. Initially the set $\mathcal{V}$ is given by all particles located inside a sphere with radius equal to $2.5 \: \rm R_e$ which is re-shaped iteratively
 using the eigenvalues of I. The distance measure is defined as  $r^2_k = x^2_k +y_k^2/q^2 +z_k^2/s^2$, where q and s are
 updated in each  iteration. The inclination angle $i$ of a galaxy is defined to be the angle between the shortest axis and the
 line-of-sight direction ($i=90^\circ$ for edge-on). 
 
Having determined a galaxy's shape, we define x, y, z axes corresponding to
the  axis lengths a, b, c ($a>b>c$) and calculate the true $\beta_z$ using mass weighted velocity moments(see {Section~\ref{jeans_model}}). 

For all galaxies, we use the bootstrap method to estimate the true errors for every parameter. They are found to be small relative to the JAM errors and thus do not have a significant effect on our results.

\begin{figure}
\center
\includegraphics[width=\columnwidth]{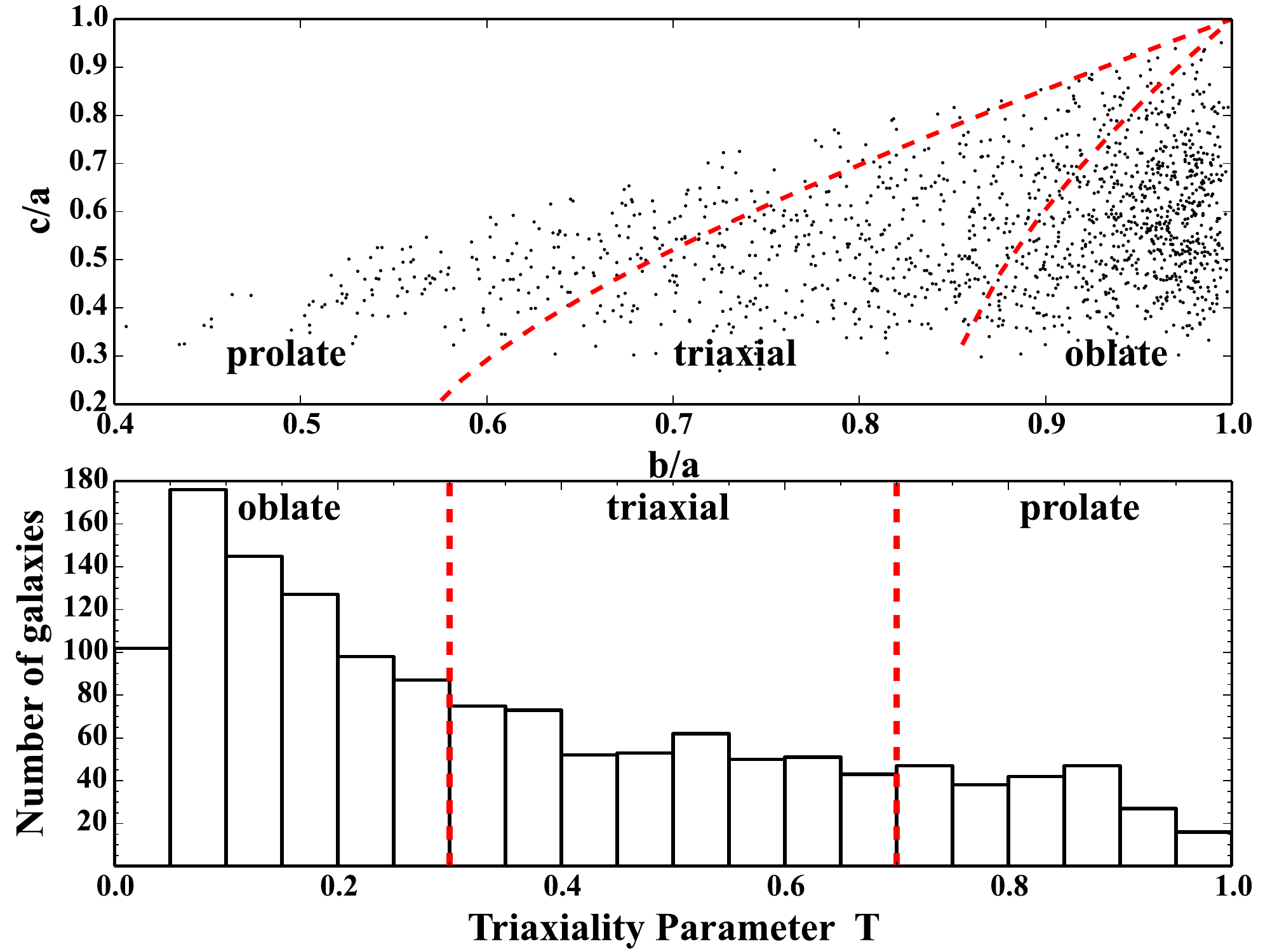}
\caption{Lower: Distribution of triaxiality parameter $T=\frac{a^2-b^2}{a^2-c^2}$.  Oblate galaxies have $T<0.3$, while prolate galaxies
have $T>0.7$. Upper: axis ratio b/a vs. c/a, the red dashed lines show the $T=0.3$ and $T=0.7$ dividing lines.}
\label{dis_tri}
\end{figure}

We use the triaxiality parameter $T\equiv \frac{a^2-b^2}{a^2-c^2}$ \citep{b8} to describe the triaxial shape of a galaxy. About $52\%$ of the
galaxies in our sample are oblate (with $T<0.3$), and about $15\%$ of the galaxies are prolate ($T>0.7$). The values of $0.3$ and $0.7$ are used to separate the oblate galaxies from the prolate galaxies.  The precise values are not too important and do not significantly alter our results.
The distribution of triaxiality parameter values for all the galaxies are shown in Fig.~\ref{dis_tri}. The shape distribution depends on the mass, and we will return to this point later in Section~\ref{sec:shape}.

In observational work, the Sersic index \citep{Sersic1963} is often used to separate early and late type galaxies.  We calculate the Sersic index $n_{\rm Sersic}$ 
for our Illustris galaxies as follows.  For each star particle in the galaxy, we generate its spectrum using the stellar population synthesis (SPS) method implemented by \citet{Bruzual2003} with a \citet{Chabrier2003} initial mass function. The spectrum can be used to 
calculate the brightness of stars in a given observational band and the brightness distribution of a galaxy. The Sersic index  is 
obtained by fitting  each galaxy's circularly averaged projected brightness profile.  Note that, the brightness distribution produced with the
SPS model is only used to estimate a galaxy's Sersic index. In our JAM modelling tests, a galaxy's brightness distribution is calculated from 
the stellar mass distribution by setting $M^*/L=1$ (as described in the second paragraph of this section).

\begin{figure}
\center
\includegraphics[width=\columnwidth]{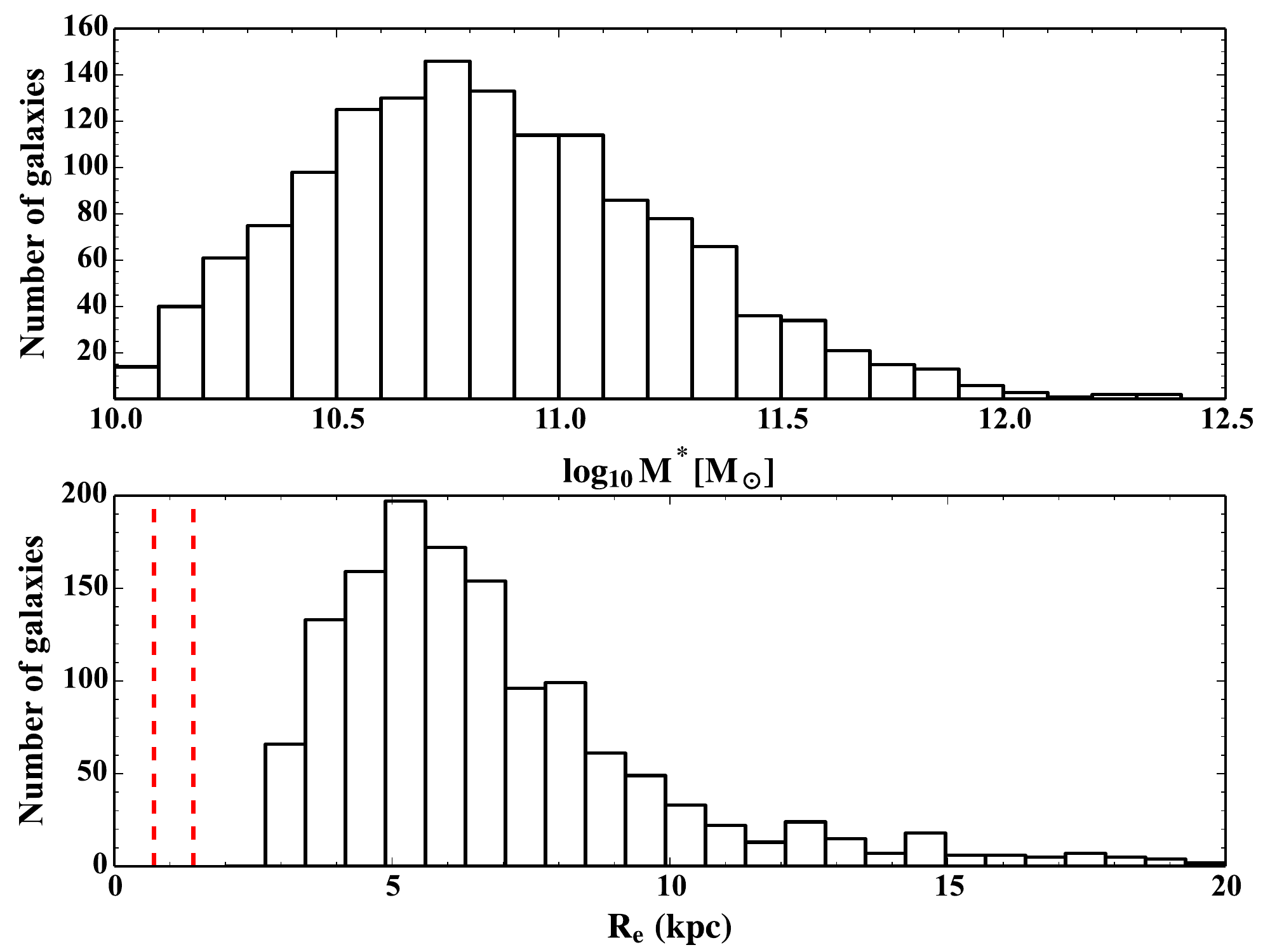}
\caption{Stellar mass (upper) and effective radius (lower) distribution of our mock galaxy sample. The red dashed lines show the softening length for star ($710\:  \rm pc$) and dark matter ($1420\:  \rm pc$) particles in the simulation. In total, there are 1413 galaxies.}
\label{stellar_mass}
\end{figure}

For our simulated galaxy sample, we select all galaxies with $M^*>10^{10} M_{\odot}$ (to avoid the large data noise inherent in small galaxies and  to match the expected MaNGA population), and with $n_{\rm Sersic}>1.5$ to include only early type galaxies.
We visually check the surface brightness density distribution of every galaxy and remove galaxies with obvious signs of being in the process of merging, or with large noise in their surface density map, which may cause failures in the MGE fitting process.
In total, our final sample contains 1413 galaxies. In the Illustris simulation, every dark halo has a unique halo ID. The galaxies in our sample are named with their dark halo IDs (e.g. subhalo12).   The galaxy stellar mass histogram is shown in Fig.~\ref{stellar_mass}. It peaks around $M^*\approx 10^{10.8}M_{\odot}$ ($\sim$ Milky Way), and tails off at $M^*\approx 10^{12}M_{\odot}$.

\begin{figure}
\center
\includegraphics[width=\columnwidth]{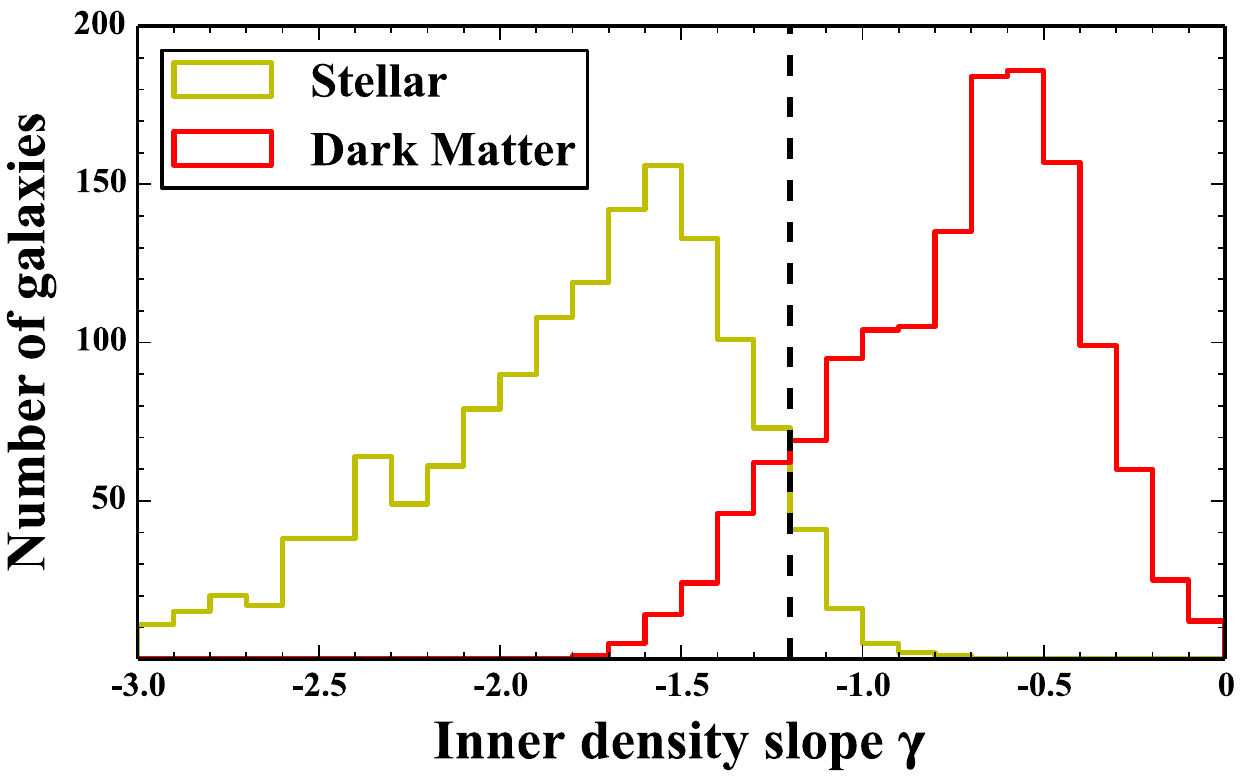}
\caption{Inner density slope distribution for dark matter (red) and stars (yellow).  The dark halo slopes are the parameter $\gamma$ obtained by fitting the true halo density profile with a gNFW model. The stellar mass density slopes are determined by fitting the true stellar density profile with $\rho^{*}\propto r^{\gamma}$ within $1 \: \rm R_e$. }
\label{slope_distribution}
\end{figure}

In Fig.~\ref{slope_distribution}, we show the distribution of the inner density slope for the dark matter and stars in our sample. It can be seen that the stellar inner slopes peak at -1.6, which is steeper than the dark matter slopes (peak at -0.7).  Most stellar inner slopes are less than -1.2 while most dark matter inner slopes are larger than -1.2. This observation is used later to set a prior in the MCMC simulation with the objective of  reducing the degeneracy between dark matter and stellar mass.

\section{Jeans-Anisotropic-MGE}\label{jam_model}
In this section, we describe the JAM + MCMC method that we use to model the galaxy stellar mass, the dark matter mass, and the total combined mass, as well as  the galaxy kinematics. We use JAM to solve the Jeans equations, and extract the best model galaxy parameters using an MCMC technique. We compare the masses and parameters so obtained with their true values to assess the effectiveness or otherwise of JAM.
\subsection{Mass model}\label{mass_model}

The mass model of a galaxy consists of two components: the stellar mass and the dark matter halo. 
Following \citet{b6}, we express both components as a set of elliptical Gaussian distributions
using the Multi-Gaussian Expansion (MGE) method described in \citet{b1} and \citet{b2}. The method is detailed below.

\subsubsection{Multi-Gaussian Expansion (MGE) method}

Following \citet{b1}, the MGE formalism is adopted to parametrize the mass model. The MGE technique 
expresses the surface brightness (density) as a summation of a series of elliptical Gaussian functions.
The galaxy surface brightness density $\Sigma(x',y')$ can be written as
\begin{equation}\label{eq:surf}
\Sigma(x',y') = \sum_{k=1}^N
{\frac{L_k}{2\pi\sigma^2_k q_k'} \exp
\left[
    -\frac{1}{2\sigma^2_k}
    \left(x'^2 + \frac{y'^2}{q'^2_k} \right)
\right]},
\end{equation}
where $L_k$ is the total luminosity of the $k$th Gaussian component with dispersion $\sigma_k$ along the major axis,
N is the number of the adopted Gaussian components, and $q'_k$ is the projected axial ratio with $0\le q'_k\le1$.
The MGE parametrization is the first and crucial step of JAM modelling. We use the {\bf mge\_fit\_sectors}\footnote{Available from http://www-astro.physics.ox.ac.uk/$\sim$mxc/software} software \citep{b2} to perform our MGEs, and follow closely the procedure described in the paper.

For each galaxy, we define its effective radius $\rm R_e$ to be the half-light radius,
calculated using the method described in \citet{b5}, from the best-fitting MGE brightness model.

We fit  MGE models to a mock galaxy's surface brightness map. Once we find the best MGE model for the surface brightness distribution, we  de-project it using the MGE formalism to obtain the 3-dimensional intrinsic luminosity density
$\nu$. The solution of the de-projection is usually non-unique, except for the edge-on axis-symmetric case \citep{Pohlen2007}. For low inclination angle
galaxies, the degeneracy is severe \citep{b3}. In the MGE formalism, a unique solution is easily obtainable if the inclination is known and  by making an assumption as to whether the galaxy is oblate, prolate or triaxial. 
Following \citet{M1992},  we de-project the surface brightness under an oblate axisymmetric assumption, the consequences of which are  discussed in Section~\ref{sec:shape}. The luminosity density $\nu$  can then be written in cylindrical polar coordinates $(R, z)$ as  
\begin{equation}\label{eq:dens}
\nu(R,z) = \sum_{k=1}^N
\frac{L_k}{(\sqrt{2\pi}\, \sigma_k)^3 q_k} \exp
\left[
    -\frac{1}{2\sigma_k^2}
    \left(R^2+\frac{z^2}{q_k^2} \right)
\right],
\end{equation}
where $L_k$ and  $\sigma_k$ for every Gaussian component are the same as in the projected case,  and  the 3-dimensional intrinsic axial ratio $q_k$ of each Gaussian component can be related
to the projected $q'_k$ by
\begin{equation}\label{eq:qmge}
    q_k=\frac{\sqrt{q'^2_k-\cos^2 i}}{\sin i},
\end{equation}
where $i$ is the galaxy inclination angle ($i$=$90^\circ$ for edge on). 

The luminosity distribution can be converted to the stellar mass distribution using the galaxy's stellar mass-to-light ratio, $M^*/L$. In this paper, we only consider constant ratios.  

Although we can obtain a good surface brightness fitting  through the MGE method (mean error $\sim 5\%$), the stellar mass density profile we obtain is usually lower than the true value at the centre of the galaxy because of finite resolution smoothing. We explore this effect by comparing the results from two different image resolutions (see {Section~\ref{rst}}).

Our mass models also include a spherical dark matter halo.
From analysing  cold dark matter simulations, haloes can be approximated by a universal mass density profile with an inner slope $\gamma=1$ and an outer
slope $\alpha=3$ \citep{b4}. 

The situation, however, becomes more complex when  baryonic processes are considered. It has been shown in hydrodynamical simulations that baryonic processes modify the inner profile of the dark halo \citep{Abadi2010}.
Following \citet{b5}, we use a generalized NFW (gNFW) dark halo
\begin{equation}
        \rho_{\rm DM}(r)=\rho_s \left(\frac{r}{R_s}\right)^\gamma
            \left(\frac{1}{2}+\frac{1}{2}\frac{r}{R_s}\right)^{-\gamma-3}.
\end{equation}
The halo density $\rho_s$ at $R_s$ is parametrized by the dark matter fraction within 1 effective radius, $f_{\rm DM}(r=R_e)$.
For computational efficiency, given $\gamma$, $R_s$ and $f_{\rm DM}$, we express the gNFW dark matter profile with  Gaussian functions using the MGE method. Thus, the total mass density profile is expressed as a series of elliptical Gaussian functions.

\subsection{Jeans Equations}\label{jeans_model} 
A detailed description of JAM modelling  is provided in \citet{b6}. Below, 
we  give a brief introduction.  A steady-state axisymmetric stellar system satisfies
two Jeans equations in cylindrical coordinates $(R, z, \phi)$ \citep{b6}
\begin{eqnarray}
    \frac{\nu\overline{v_R^2}-\nu\overline{v_\phi^2}}{R}
    + \frac{\partial(\nu\overline{v_R^2})}{\partial R}
    + \frac{\partial(\nu\overline{v_R v_z})}{\partial z}
    & = & -\nu\frac{\partial\Phi_{\rm tot}}{\partial R},
    \label{eq:jeans_cyl_R}\\
    \frac{\nu\overline{v_R v_z}}{R}
    + \frac{\partial(\nu\overline{v_z^2})}{\partial z}
    + \frac{\partial(\nu\overline{v_R v_z})}{\partial R}
    & = & -\nu\frac{\partial\Phi_{\rm tot}}{\partial z},
    \label{eq:jeans_cyl_z}
\end{eqnarray}
where 
\begin{equation}
    \nu\overline{v_k v_j}\equiv\int v_k v_j f\; \mathrm{d}^3 \mathbf{v},
\end{equation}
and $f$ is the distribution function of the stars, $\Phi_{\rm tot}$ is the gravitational potential, $\nu$ is the luminosity density (see Eq.~\ref{eq:dens}).

Our mass model consists of two components, the stellar distribution and the dark matter halo, both of which we represent by a sum of Gaussian functions. 
The potential generated by these density distributions is given by \citep{b1}
\begin{equation}
   \Phi_{\rm tot} (R,z)=  -\sqrt{2/\pi}\, G \int_0^1 \sum_{j=1}^K{\frac{M_j\, {\mathcal H}_j(u)}{\sigma_j}} \mathrm{d} u,
\end{equation}
where $G$ is the the gravitational constant, $M_j$ is the total mass of  Gaussian component $j$, the summation is over all the $K$ Gaussian functions of the  stellar and the dark matter components, and
\begin{equation}
\mathcal{H}_j(u) = \frac{{\exp \left\{ - \frac{{u^2 }}
{{2\sigma _j^2 }}\left[ {R^2  + \frac{{z^2 }}{{1 - (1 - q_j^2 )u^2 }}}
\right] \right\}}}{{\sqrt {1 - (1 - q_j^2 )u^2 } }}.
\label{eq:integral}
\end{equation}

Given $\Phi_{\rm tot}$ and $\nu$, Equations~(\ref{eq:jeans_cyl_R}) and
(\ref{eq:jeans_cyl_z}) depend on four unknown quantities $\overline{v_R^2}$,
$\overline{v_z^2}$, $\overline{v_\phi^2}$ and $\overline{v_R v_z}$, 
and therefore additional assumptions are required to determine a
unique solution. One common choice is to align the orientation 
 of the velocity dispersion ellipsoid with the meridional plane $(R,z)$ and then set the shape within that plane. 

In this paper,  we follow the assumptions made in \citet{b6}: (i) the velocity
dispersion ellipsoid is aligned with the cylindrical coordinate system ($\overline{v_R v_z}=0$) and (ii) the
anisotropy in the meridional plane is constant, i.e. $\overline{v_R^2} = b\overline{v_z^2}$. 
Similarly to the standard definition of the anisotropy parameter \citep{b8}, the anisotropy parameter in the $z$ direction $\beta_{z}$ can be written as
\begin{equation}
  \label{eq:beta}
  \beta_{z} \equiv 1 - \frac{\overline{v_z^2}}{\overline{v_R^2}} \equiv 1 - \frac{1}{b} \, .
\end{equation}

The Jeans equations  thus reduce to
\begin{eqnarray}
    \frac{b\,\nu\overline{v_z^2}-\nu\overline{v_\phi^2}}{R}
    + \frac{\partial(b\,\nu\overline{v_z^2})}{\partial R}
    & = & -\nu\frac{\partial\Phi_{\rm tot}}{\partial R},
    \label{eq:jeans_beta_R}\\
    \frac{\partial(\nu\overline{v_z^2})}{\partial z}
    & = & -\nu\frac{\partial\Phi_{\rm tot}}{\partial z}.
    \label{eq:jeans_beta_z}
\end{eqnarray}

If we set the boundary condition  $\nu\overline{v_z^2}=0$ as $z\rightarrow\infty$,
we can write the solution as
\begin{eqnarray}
    \nu\overline{v_z^2}(R,z)
    & = & \int_z^\infty \nu\frac{\partial\Phi_{\rm tot}}{\partial z}{\mathrm{d}} z
    \label{eq:jeans_sol_z}\\
    \nu\overline{v_\phi^2}(R,z) & = &
    b\left[
    R \frac{\partial(\nu\overline{v_z^2})}{\partial R}
    + \nu\overline{v_z^2} \right]
    + R \nu\frac{\partial\Phi_{\rm tot}}{\partial R}
    \label{eq:jeans_sol_R}.
\end{eqnarray}
These intrinsic quantities must then be integrated along the line-of-sight to obtain the projected second velocity moment $\overline{v^{2}_{\mathrm{los}}}$,
which can be directly compared with the stellar kinematic observables, i.e. the root mean square velocity $v_{\mathrm{rms}} \equiv \sqrt{v^{2} + \sigma^{2}}$, where $v$
and $\sigma$ are the line-of-sight stellar mass weighted mean velocity and velocity dispersion, respectively.

Recall that the assumptions we choose to perform JAM modelling with are:
\begin{enumerate}
\item an oblate shape for the luminosity density distribution; 
\item{a constant stellar mass to light ratio in all radii};
\item a constant anisotropy in the meridional plane, and
\item a double power law  dark matter profile (the gNFW halo profile). 
\end{enumerate}

By assessing the extent to which the model predicted $v_{\rm rms}$ matches the mock galaxy's $v_{\rm rms}$, we are able to estimate  the following six parameters: 
\begin{enumerate}
\item{the inclination $i$ (the angle between the line of sight and the axis of symmetry)};
\item the anisotropy parameter $\beta_z$ in Equation~(\ref{eq:beta});
\item the stellar mass-to-light ratio, $M^*/L$;
\item and the three parameters in the dark matter halo: $f_{\rm DM}(r=R_e)$, $\gamma$ and $r_{s}$. 
\end{enumerate}

For the actual modelling, we use the {\bf jam\_axi\_rms}\footnote{Available from http://www-astro.physics.ox.ac.uk/$\sim$mxc/software}  software.
 
\subsection{Bayesian Inference and the MCMC method}
\label{mcmc}

We adopt the Markov Chain Monte Carlo (MCMC) technique for model inferences.

If we denote the parameter set as $\bmath{p}$ and the data set as $\bmath{d}$,
from Bayes' theorem, the posterior probability distribution function (PDF) for the set of parameters~$\bmath{p}$ can be written as
\begin{equation}
  \label{eq:posterior}
  \mathrm{P}(\bmath{p} \, | \, \bmath{d}) =
  \frac{
  \mathrm{P}(\bmath{d} \, | \, \bmath{p}) \, 
  \mathrm{P}(\bmath{p})
  }
  {\mathrm{P}(\bmath{d})} ,
\end{equation}
where $\mathrm{P}(\bmath{d} \, | \, \bmath{p})$ is the likelihood, $\mathrm{P}(\bmath{p})$ is the prior, and
$\mathrm{P}(\bmath{d})$ is the factor required to normalize the posterior over~$\bmath{p}$, which is valuable
for model selection.

The set of parameters~$\bmath{p}_{\mathrm{MAX}}$ for which the
posterior probability is maximized is interpreted as our `best' model since it is the parameter set that is able to reproduce the data most closely.    

A good way to find out the $\bmath{p}_{\mathrm{MAX}}$ is through the Markov Chain Monte Carlo (MCMC) method. With this method, we can sample $\bmath{p}$ in parameter space according to the joint posterior probability $\mathrm{P}(\bmath{p} \, | \, \bmath{d}) $. Once we characterise the posterior probability by these samples, the marginalized posterior PDFs for individual parameter $\bmath{p}_{i}$ are approximated by the histograms of these samples. There are several ways to draw $\bmath{p}_{\mathrm{MAX}}$ from these one-dimension histograms, e.g., using mean, maximum or median. In this paper, we choose the medians of the one-dimension posterior distributions as the best-fitting parameters while the parameter uncertainties are calculated with their 68\% confidence intervals by taking the 16th and 84th percentiles. The six parameters sampled by MCMC are $(i, \beta_z, f_{\rm DM}, \gamma, R_s, M^*/L)$. 

Assuming the observational errors are Gaussian, we have
\begin{equation}
    \mathrm{P}(\bmath{d} \, | \, \bmath{p})\propto \exp\left(-\frac{\chi^2}{2}\right),
\end{equation}
with
\begin{equation}
    \chi^2 = \sum_j \left(\frac{ \langle v^2_{\rm los}\rangle_j^{1/2} - v_{{\rm rms},j} }{ \Delta v_{{\rm rms},j} }\right)^2,
\end{equation}
where ${ \Delta v_{{\rm rms},j} }$ is set to be the Poisson error ($1 / \sqrt{N_j}$, where $N_j$ is the number of the particles in grid cell $j$) scaled to be between $5 \: \rm{km\: s^{-1}}$ and $40 \: \rm{km\: s^{-1}}$,  which is the typical error range of a high signal-to-noise ratio CALIFA galaxy.  Since the spectral resolution of the MaNGA survey is higher than that of CALIFA, the error for MaNGA galaxies should be somewhat smaller. The sum is taken over all the cells $j$  within a $2.5\: \rm R_e$ radius on the 2D kinematic grid.

In this paper, we fit $v_{\rm rms}$ for the simulated galaxies out to 2.5$ \: \rm R_e$;

In order to be consistent with \cite{b5}, we set flat priors within given bounds:
\begin{enumerate}
\item According to Fig.~\ref{slope_distribution}, we set a lower boundary of -1.2 for the dark matter inner slope $\gamma$;
\item $f_{\rm DM}$ is between $0$ and $1$;
\item $M^*/L$ is between 0 and 10;
\item $R_s$ is between $15$ and $40$  {\rm kpc} (corresponds to the scaling radius of the gNFW-fitted dark matter haloes in our sample);
\item Following \cite{b5}, we limit $\beta_z$  to the range $[0,0.4]$;
\item $\cos i$ is between $0$ and $\cos i_{\rm low}$, where $i_{\rm low}$ is the  minimum inclination derived from Equation~(\ref{eq:qmge}), in order to obtain a physical intrinsic axis-ratio $q_k$. 
\end{enumerate}

\begin{figure*}
\center
\includegraphics[width=\textwidth]{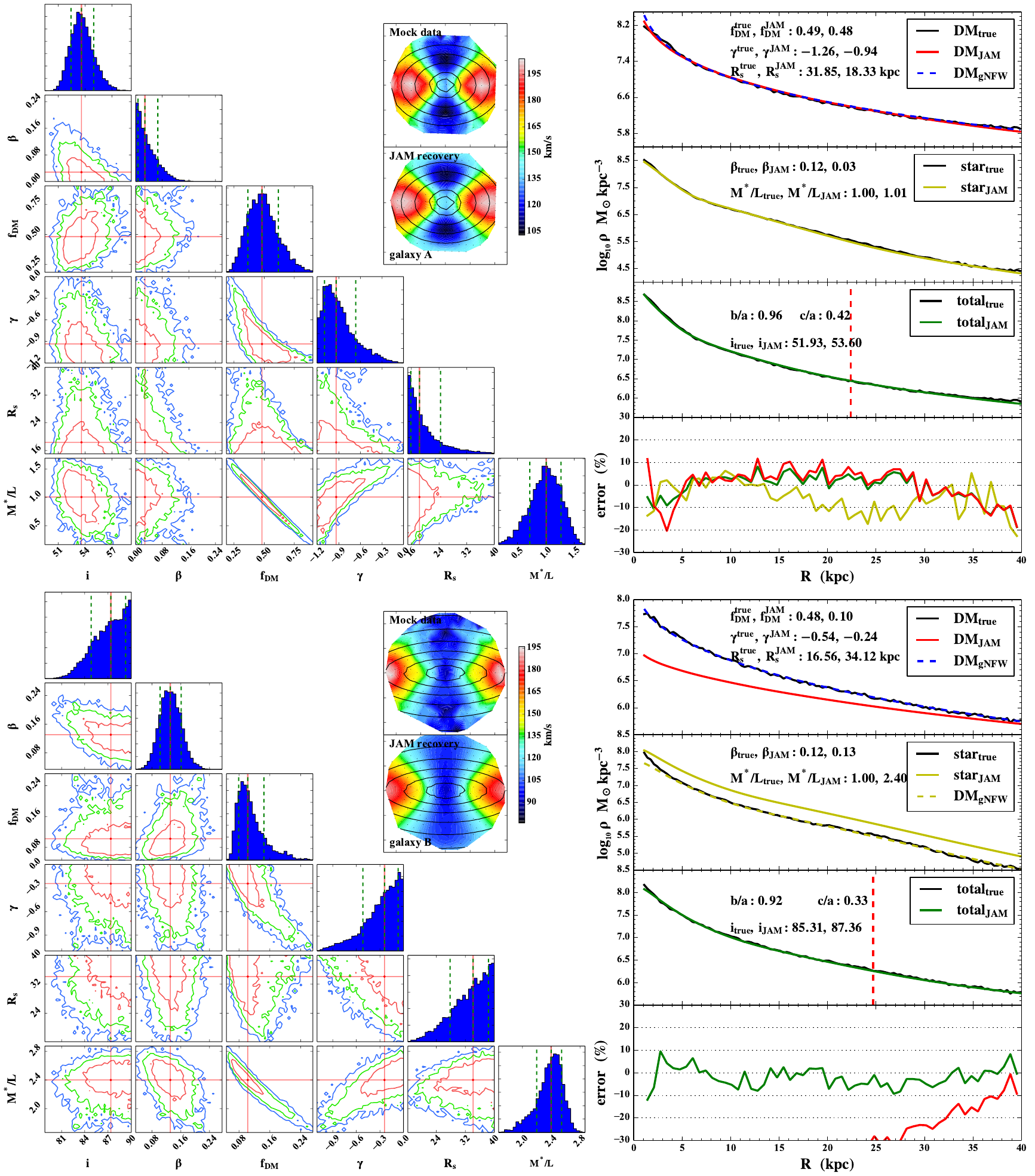}
\caption{ JAM modelling for two simulated galaxies: galaxy A (top) and galaxy B (bottom). The contours on the left side
show the 2D marginalized distributions. The red, green and blue contours are 1, 2 and 3$\sigma$ confidence level respectively. 
The red cross in each plot indicates the value we choose for the best model parameter set.
The blue histogram in every row shows the 1D marginalized posterior distribution, with red lines marking the median and the green lines,  the $1\sigma$ region.
In the right hand panels,  the subplots  on the upper right  show the root-mean-squared velocity
 $(v_{\rm rms}\equiv \sqrt{v^{2}+\sigma^{2}})$ map, where $v$ is the mean stellar velocity
and $\sigma$ is the stellar velocity dispersion. Ticks are separated by $10 \ \rm kpc$. 
On the right side, we compare the input and fitted density profiles. Top: dark matter density profile (black for true, red for JAM prediction and the blue dashed line is the  best-fitting gNFW profile for the input density field). Middle:  stellar mass density profile (black for true, yellow solid for the JAM prediction, and yellow dashed is the density profile when $M^*/L=1$. 
Bottom:  the total mass density profile (black for true and green for JAM prediction). The red vertical dashed line shows the kinematic data range that we use ($2.5 \: \rm R_e$). Bottom: the relative error for each profile (red for the dark matter, yellow for stellar mass, green for the total).
The true parameter values for the mock galaxies and their best-fitting JAM values are given in each subplot. 
\label{good_fit}}
\end{figure*}

We sample the  posterior distribution of parameters  with the  ``emcee" code  from \citet{b9},
 which provides a fast and stable implementation of an affine-invariant ensemble sampler for MCMC
  which  can be parallelized without extra effort.


\section{Results}\label{rst}
Before we study the mock galaxies, we first checked the functionality of the JAM method using SAURON galaxies, and verified that we obtained nearly the same results with \citet{b6}. 

As mentioned before, in order to investigate the effects of image resolution, we generate  luminosity distributions 
with two different image resolutions:  the low resolution image ($L_{\rm cell}=2 \rm kpc/h$) and
the high resolution image ($L_{\rm cell}=0.5 \rm kpc/h$). The redshift range of MaNGA survey is $0.01-0.15$ \citep{b11}, and at a redshift of $0.1$ these sizes correspond to 1.52 arcsec and 0.38 arcsec respectively, roughly the SDSS seeing and the best seeing from the ground.

In Section~\ref{sec:individual}, we show  the JAM modelling results of two individual mock galaxies.  Statistical results for our simulated galaxy sample are given in Section~\ref{sec:statistical}.

\subsection{Results of two individual galaxies}\label{sec:individual}

In Fig.~\ref{good_fit}, we show the JAM modelling  results for two mock galaxies, denoted as galaxy A (subhalo224435) and galaxy B (subhalo362540), selected for their well fitted $v_{\rm rms}$ maps. 
The results are for high resolution images. Both galaxies are oblate, with axis ratio $b/a>0.9$. 
We show the input and the best-fitting $v_{\rm rms}$ in the subplots and,
as can be seen, JAM does indeed successfully recover the $v_{\rm rms}$ map of each galaxy. 

\begin{figure*}
\center
\includegraphics[width=0.8\textwidth]{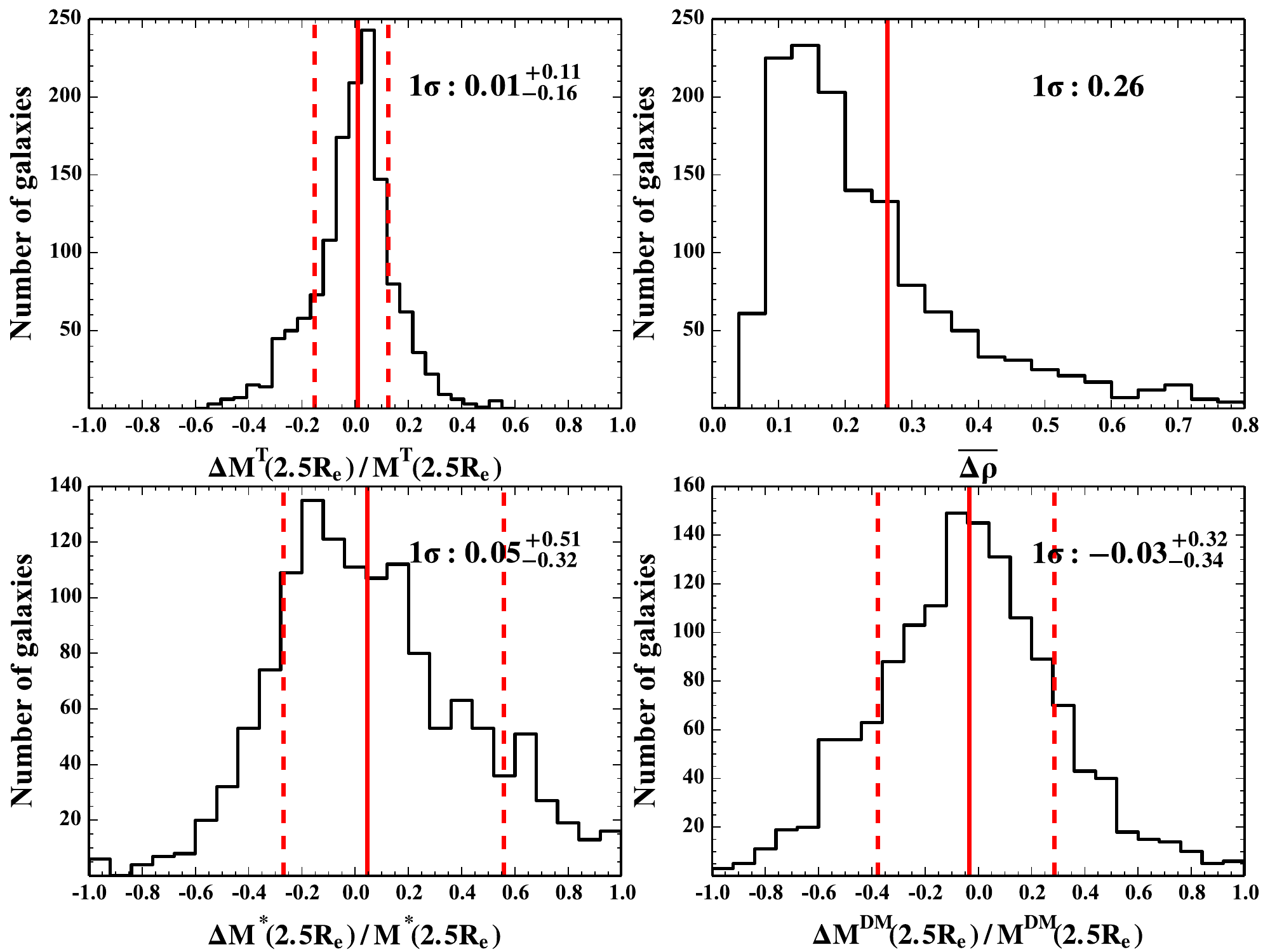}
\caption{Histogram of the fractional error in mass for  the 1413 simulated galaxies with high-resolution images. Upper left: fractional error of the enclosed total mass $\Delta M^{\rm T}(2.5\: \rm R_e)/M^{\rm T}(2.5 \: \rm R_e)$. Upper right: mean fractional error of the total mass density profile $\overline{\Delta \rho}$. Lower left: fractional error of the enclosed stellar  mass $\Delta M^*(2.5\: \rm R_e)/M^*(2.5\: \rm R_e)$.  Lower right: fractional error of the enclosed dark mass $\Delta M^{\rm DM}(2.5\: \rm R_e)/M^{\rm DM}(2.5\: \rm R_e)$. The red vertical solid and dashed lines show the median and $1\sigma$ regions for enclosed mass error, and $1\sigma$ error only for the mean density error. Their values are listed on the top right of each subplot. 
\label{accuracy}}
\end{figure*}

In the upper panels, we show the results for galaxy A. The corner-plots on the left side show the posterior
distribution of model parameters. On the right side, we compare the JAM fitted galaxy density profile (spherically averaged) 
with the data input. It can be seen that the JAM method not only fits the total density profile well, but also reproduces
the density profile of dark matter and stellar mass with mean density profile error less than $10\%$. The best-fitting model parameters and their true values (see Section~\ref{sec:mock} for the calculation of the true parameters) are listed in the panels.
The input galaxy parameters ($M^*/L$, $f_{\rm DM}$, $\gamma$, $R_s$) are $(1.0, 0.49, -1.26, 31.85)$, and the JAM best-fitting
parameters are $(1.01_{-0.289}^{+0.260}, 0.48_{-0.12}^{+0.14}, -0.94_{-0.16}^{+0.27}, 18.33_{-2.5}^{+6.1})$. While the anisotropic parameter
$\beta_z$ is 0.03, different from the true value 0.12, the accuracy of $M^*/L$ and  $f_{\rm DM}$ recovery is within 2\%.
The 2-D posterior plots show a strong degeneracy between $f_{\rm DM}$ and $M^*/L$. A degeneracy is also
present between $\gamma$ and $f_{\rm DM}$.

For galaxy B, the total density profile is also well recovered. However, the best-fitting  $f_{\rm DM}$ and $M^*/L$ deviate significantly from the true values. The input galaxy parameters ($M^*/L$, $f_{\rm DM}$, $\gamma$, $R_s$) are $(1.0, 0.48, -0.54, 16.56)$, and the JAM best-fitting
parameters are $(2.40_{-0.203}^{+0.145}, 0.10_{-0.02}^{+0.04}, -0.24_{-0.27}^{+0.17}, 34.12_{-6.3}^{+4.3})$. It can be seen that the model prediction of $f_{\rm DM}$ is $0.1$, while the true value is $0.48$. The best-fitting $M^*/L$ is $2.4$, more
than two times larger than the input value. It can be seen that both the dark matter profile and the stellar mass profile deviate from the true ones. On the other hand, 
 JAM still reproduces well the total density profile and the total mass with an accuracy of 10\%. The bias in $f_{\rm DM}$ and $M^*/L$ is caused by the limited image resolution used in MGE fitting (see {Section~\ref{mass_model}}). From the second subplot on the right panel, it can be seen that the inner slope of the stellar mass profile is not well
reproduced. In the JAM method, the shape of the stellar mass profile is determined by MGE fitting to the surface brightness distribution of the galaxy.
In the MCMC process,  only the amplitude of the stellar mass profile is allowed to change by varying the $M^*/L$.
As a result, the underestimation of the inner slope of the stellar mass translates to an overestimation of the  $M^*/L$ 
and an underestimation of the dark matter fraction, $f_{\rm DM}$.

\begin{figure}
\center
\includegraphics[width=\columnwidth]{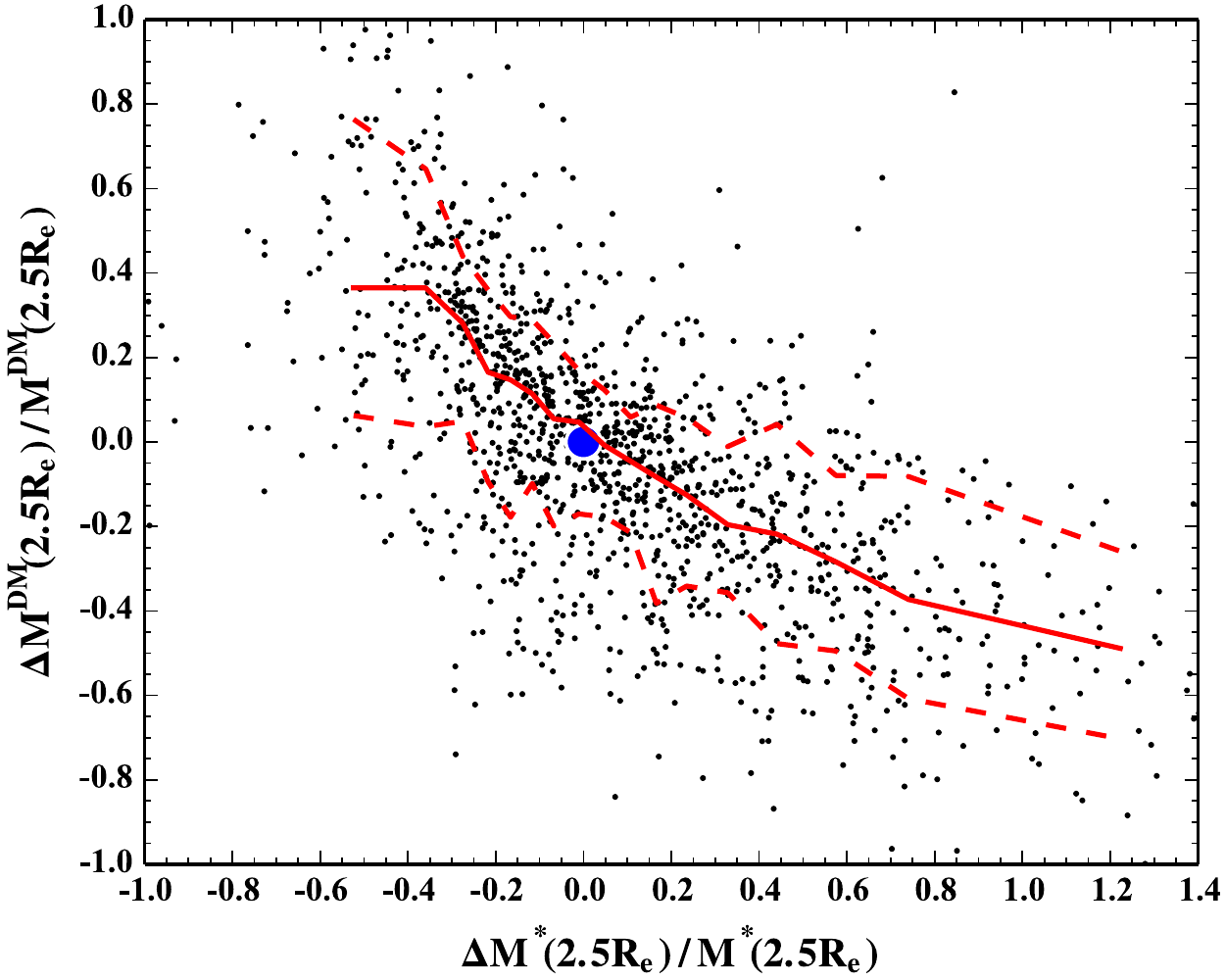}
\caption{Fractional error of the stellar mass $\Delta M^*(2.5\: \rm R_e)/M^*(2.5\: \rm R_e)$ vs. the fractional error of the dark matter mass $\Delta M^{\rm DM}(2.5\: \rm R_e)/M^{\rm DM}(2.5\: \rm R_e)$ for the 1413 simulated galaxies with high-resolution images. The red solid line is the median and the red dashed lines show the $1\sigma$ region in each bin. The blue circle shows the zero error point. 
\label{degeneracy_statistical}}
\end{figure}

\subsection{Statistical results}\label{sec:statistical}

\subsubsection{Accuracy of the mass model}
In this subsection, we discuss the accuracy and the bias of the mass model predicted by the JAM method. 

To quantify the accuracy of mass recovery, we define the mass estimation error within radius $R$ as
\begin{equation}
\label{eq:perror}
\frac{\Delta M(R)}{M(R)}=\frac{M(R)_{\rm JAM}-M(R)_{\rm True}}{M(R)_{\rm True}},
\end{equation}
where $M(R)_{\rm JAM}$ is the mass enclosed in a sphere within radius R predicted by the JAM method, and  $M(R)_{\rm True}$ is the true mass of the simulated galaxy.

We also define the mean error of the density profile as
\begin{equation}
\label{eq:terror}
\overline{\Delta \rho}=\frac{1}{R} \int _0^R \frac{\mid \rho(r)_{\rm JAM}-\rho(r)_{\rm True}\mid}{\rho(r)_{\rm True}}\mathrm{d}r,
\end{equation}
where $\rho(r)_{\rm JAM}=\rho(r)_{\rm gNFW}+\frac{M^*}{L}\overline{\nu (r)}$ is the spherically averaged total mass density profile, calculated by using the JAM best-fitting parameters, and $\rho(r)_{\rm True}$ is the true density profile of the input galaxy. 
We calculate both $\overline{\Delta \rho}$ and $\Delta M(R)$ within $R=2.5\: \rm R_e$.

We first show the results for all mock galaxies in the high resolution case. Fig.~\ref{accuracy} shows the histograms of the fractional errors of the enclosed total mass, stellar mass, dark matter mass and total density profile for all 1413 simulated galaxies. The accuracy of the JAM recovered total mass  reaches $11\%-16\%$.  The mean error of the total density profile is $26\%$, which means that we can obtain a good estimate of the total mass distribution. The bias in the recovered total mass  is only $1\%$. For the high resolution mock galaxies, the total mass obtained through the JAM method is not biased. 

Although we can  obtain  good recovery of the total mass, the errors in the individual stellar and dark matter masses are larger.  As shown in the lower
panel of  Fig.~\ref{accuracy}, the errors in the enclosed stellar and dark matter masses are $32\%-51\%$, much larger than that of the total mass. However the biases in the stellar and dark matter masses are quite small ($5\%$ and $-3\%$). There are several outliers with $\Delta M^*/M^*$ very close to $-1$, as shown in Fig.~\ref{accuracy}. These galaxies are massive galaxies with satellites within the kinematic data area which affect the results. The number of outliers ($\sim 10$) is much smaller than the whole sample ($1413$), so they will not have significant impact on our statistical results since we take the median and scatter around the median.

Good recovery of the total mass implies that error in the stellar mass may be correlated with that in the dark matter mass.
We plot the fractional error of the enclosed stellar mass vs. the fractional error of the enclosed dark mass in Fig.~\ref{degeneracy_statistical}.
Most of the galaxies are located in a long band from the upper left to the lower right, showing a strong 
correlation between $\Delta M^*$ and $\Delta M_{\rm DM}$.  Theoretically, $v_{\rm rms}$ of the stars  depends on the
gravitational potential $\Phi_{\rm tot}$, which is determined by both the dark matter and the stellar masses.
In the JAM method, the density of the stellar component is measured from the observed image with a MGE. Only the amplitude
of the stellar density is allowed to vary in the MCMC fitting process. Therefore, the effect of the limited image resolution on the MGE modelling can lead to 
a bias in disentangling the dark matter and the stellar mass. On the other hand, as long as we use a dark matter model, appropriately parametrised for flexibility, the total density profile and the total mass can be well constrained.

\begin{figure}
\center
\includegraphics[width=\columnwidth]{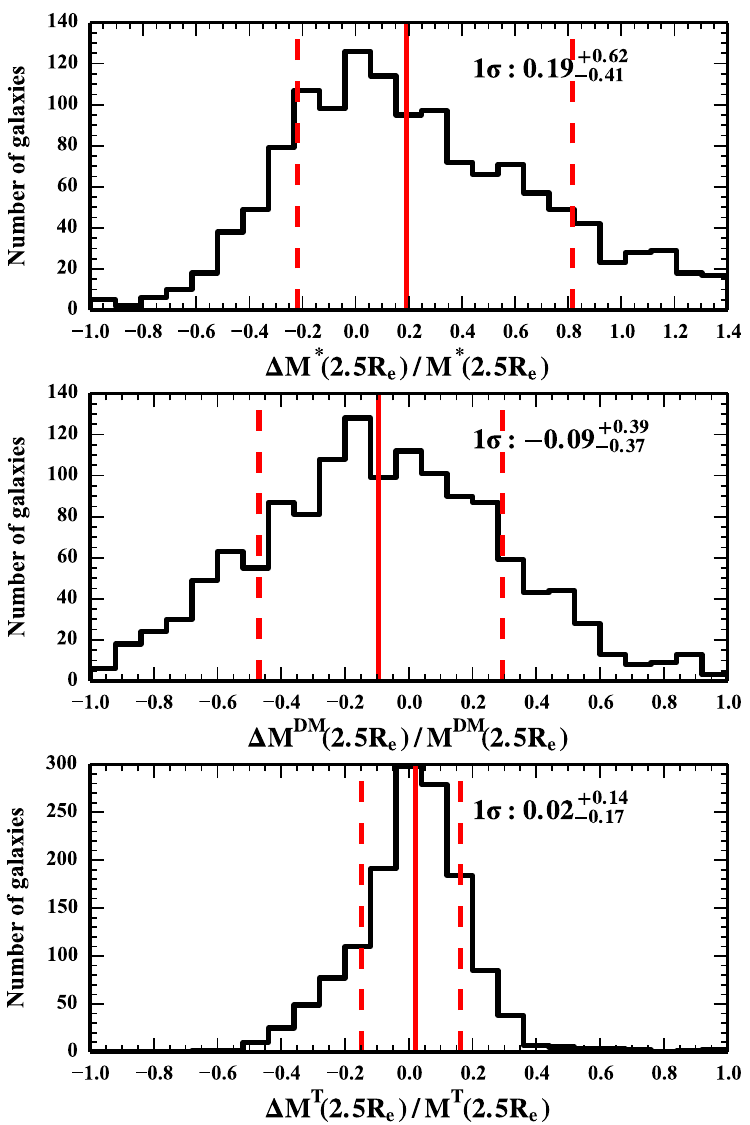}
\caption{Histogram of the mass error for the 1413 simulated galaxies with low-resolution images. Upper: enclosed stellar mass error $\Delta M^*(2.5\: \rm R_e)/M^*(2.5\: \rm R_e)$. Middle: enclosed dark mass error $\Delta M^D(2.5\: \rm R_e)/M^D(2.5\: \rm R_e)$ Lower: enclosed total mass error $\Delta M^T(2.5\: \rm R_e)/M^T(2.5\: \rm R_e)$. The labels and legends are the same as in Fig.~\ref{accuracy}.
\label{image}}
\end{figure}

\begin{figure}
\center
\includegraphics[width=\columnwidth]{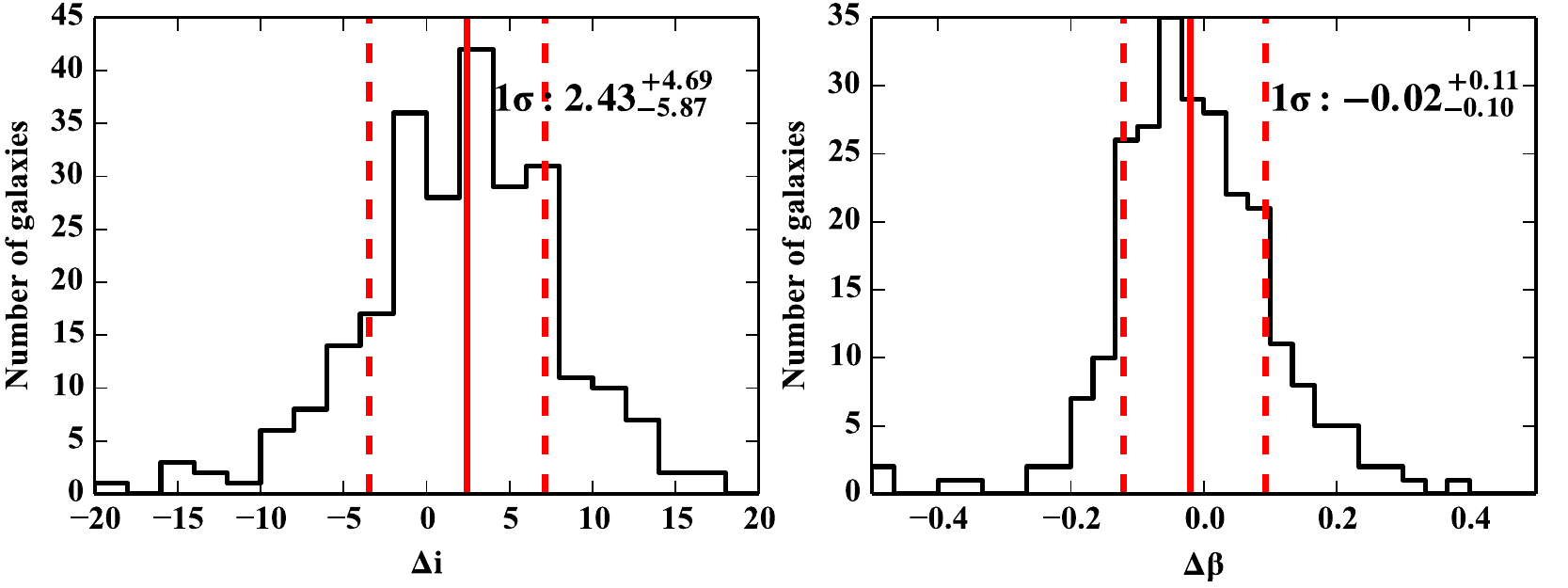}
\caption{Histogram of the inclination errors ($\Delta i$, left panel) and anisotropy errors ($\Delta \beta_z$, right panel)  for oblate galaxies with high-resolution images. The labels and legends are the same as in Fig.~\ref{accuracy}.
\label{err_inc}}
\end{figure}

In Fig.~\ref{image}, we show the results for the lower-resolution mock galaxies.   We find a significant bias with a larger scatter in
the estimated stellar mass.  In the high-resolution case, the bias is $5\%$, while in the low-resolution case, the bias is $19\%$. 
The bias in the dark matter mass estimation also increases from $-3\%$ to $-9\%$. For the low-resolution case, the MGE modelling is strongly affected in the inner regions, which may cause a larger error in the mass model.  Our results show that higher image resolutions can be effective in actually reducing the biases in the mass model.
\begin{figure*}
\center
\includegraphics[width=0.8\textwidth]{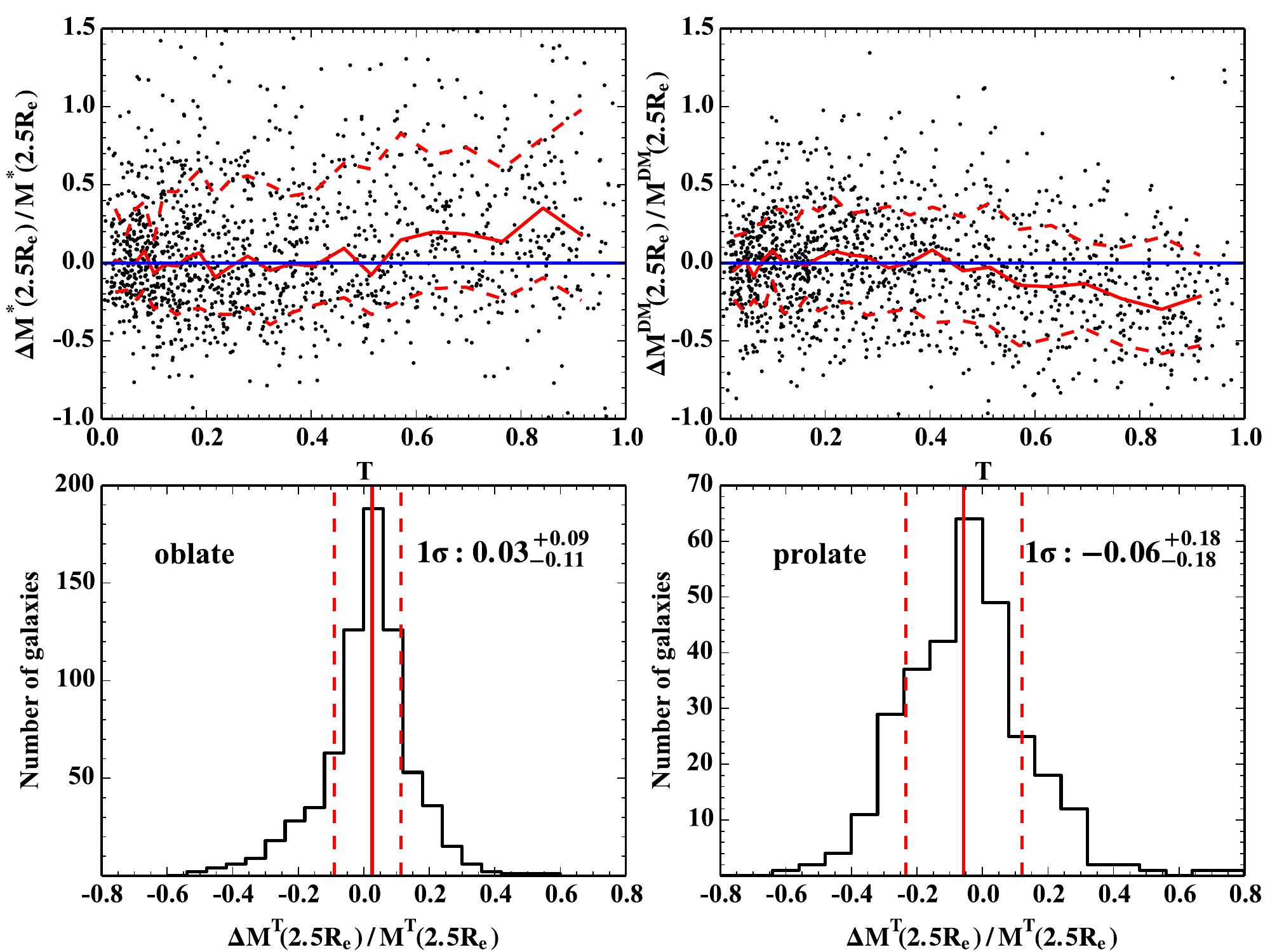}
\caption{Bias and accuracy dependence on the triaxiality parameter $T$. Upper left: triaxiality parameter $T$ vs. error in the enclosed stellar mass within $2.5\: \rm R_e$. Upper right: triaxiality parameter $T$ vs.  error in enclosed dark mass within $2.5\: \rm R_e$. The red solid line shows the median in each bin, and the red dashed lines show the  $1\sigma$ errors. The blue solid line represent an error equal to zero. Lower left: distribution of enclosed total  mass error for oblate galaxies ($T<0.3$).  Lower right: distribution of enclosed total  mass error for prolate galaxies ($T>0.7$). The red vertical solid and dashed lines show the median and $1\sigma$ region respectively, and their values are in the upper right of each panel. For prolate galaxies, the accuracy of the total mass is about $18\%$, while it is $10\%$ for oblate. In the prolate case, the predicted stellar masses and dark matter masses are also biased, the JAM method preferring more stellar mass instead of dark matter.
\label{bias_tri}}
\end{figure*}

\subsubsection{Uncertainty in the inclination and the anisotropy parameter $\beta_z$}
For prolate galaxies, the inclination, which is  defined as the angle between the shortest axis and the line-of-sight, is not meaningful since a change in the inclination  is just a rotation along the major axis. The anisotropy parameter $\beta_z$ has a similar problem, and so hereafter in this section we only discuss the results for oblate galaxies. In Fig.~\ref{err_inc}, we show the errors in the inclination and anisotropy parameters for oblate galaxies with true inclination $i_{true}>60^{\circ}$ and high resolution images. 

From Fig.~\ref{err_inc} it can be seen that the uncertainty of the JAM determined inclination is about 5 degrees with a bias of 2 degrees. The uncertainty of $\beta_z$ is 0.11 with a bias of -0.02. By comparison, \citet{b23} gives an inclination error of less than 5 degrees. However, considering our simulated galaxies are more complex, our larger error is acceptable. The  $\beta_z$ error in \citet{b23} is larger than 0.1, consistent with our results. However, in our study, when galaxies are near face on, the uncertainty of inclination increases to more than 15 degrees with a bias as large as 20 degrees. This is the consequence of the prior imposed on the inclination (see {Section~\ref{mcmc}}). This prior gives a lower limit to the inclination that can be used  in a galaxy's models. Moreover, the errors in $\beta_z$  also increase to 0.15 with a bias of -0.11.  Thus, from our study, we recover galaxy inclinations well for high inclination galaxies ($i_{true}>60^{\circ}$),  while  recovery of the velocity anisotropy from JAM models is less accurate. The recovered value of $\beta_z$ strongly depends on our model assumptions about the velocity anisotropy (constant anisotropy) and so JAM constrains this parameter less well.

\subsubsection{Dependence on the galaxy shape}\label{sec:shape}
The MGE formalism can deal with generalized geometries, including triaxial shapes. In many earlier papers, as well as in this one, a simplifying axisymmetric oblate shape is assumed. In reality, however, a large fraction of galaxies are not oblate. To investigate how the mass model is
affected when  the oblate assumption is violated, we plot in Fig.~\ref{bias_tri} the model mass error as a function of 
the triaxiality parameter for our simulated galaxies. The red solid lines represent the
median of the  mass error. It is clear that the bias in stellar mass estimation is small when $T<0.4$
and grows as $T$ increases and the galaxies become more prolate. When  $T$ approaches 1, the stellar mass predicted by JAM is $18\%$ higher than the
true value and the estimated  dark matter mass is $22\%$ lower.

In the lower panel of Fig.~\ref{bias_tri}, we show the histogram of the errors in the enclosed total mass for oblate and prolate galaxies. 
In addition to the errors in the stellar and the dark matter mass estimates, the error in the enclosed total mass is also larger 
for prolate galaxies. The scatter of the prolate galaxy mass errors is  about twice that of oblate galaxies.
The results show that  violation of the oblate assumption in the de-projection process is a key factor influencing the
accuracy of the mass model recovery.

\begin{figure}
\center
\includegraphics[width=\columnwidth]{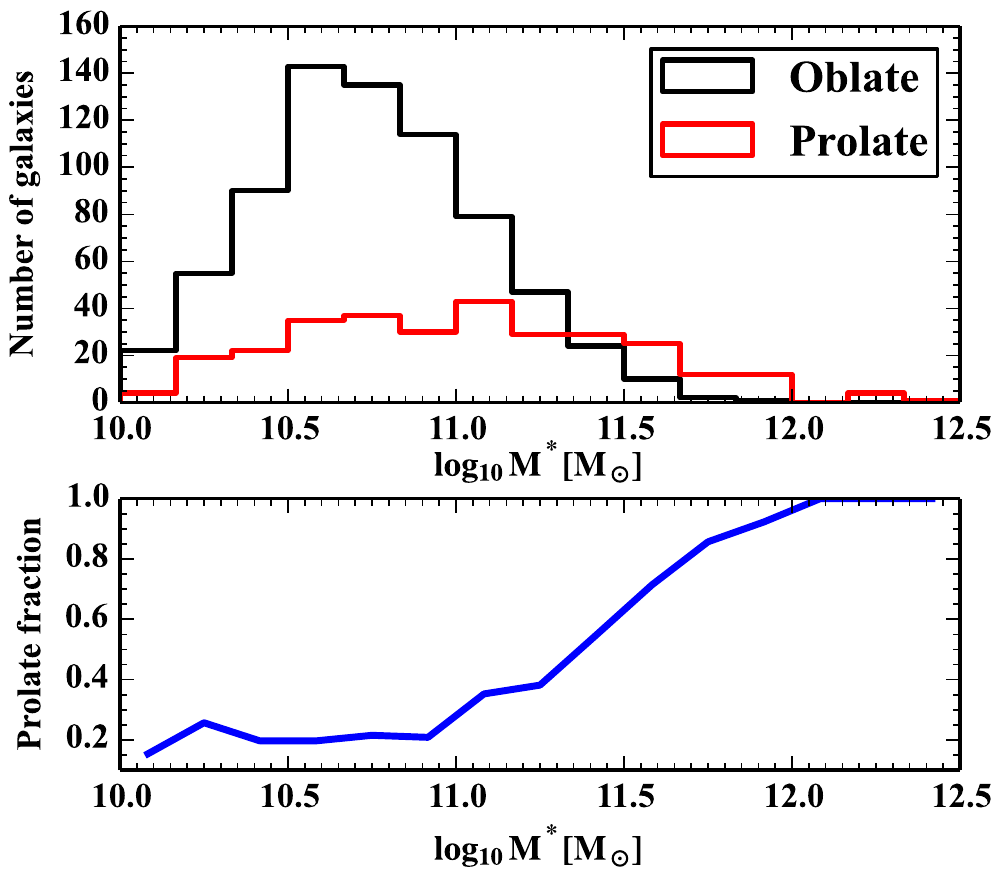}
\caption{Upper: the histogram of the stellar mass function for oblate (black) and prolate (red) galaxies.  Lower: the prolate fraction ($N_{\rm prolate} / (N_{\rm prolate} + N_{\rm oblate})$ in each mass bin) vs. the stellar mass.
\label{mass_tri}}
\end{figure}

It is interesting to note that the triaxiality parameter depends on the stellar mass (see Fig.~\ref{mass_tri}) 
with most massive galaxies being prolate, and it is this shape which  has a direct impact on mass estimation.

\subsubsection{Dependence on the galaxy rotation}
JAM was first designed to handle a set of fast-rotator galaxies in the SAURON project \citep{b6}. The assumption that the velocity ellipsoid is aligned with the cylindrical coordinate system used in JAM is a good description of the anisotropy for these fast rotators \citep{Cappellari2007}. In addition to the galaxy shape, we also check for any bias differences between fast and slow rotators. In order to quantify the amount of galaxy rotation, we use the parameter $\lambda_{\rm R}$ defined in \citet{Emsellem2007}
\begin{equation}
\lambda_{\rm R}=\frac{\sum_{i=1}^{N}M_iR_i{|V_i|}}{\sum_{i=1}^{N}M_iR_i\sqrt {V_i^2+\sigma_i^2}},
\end{equation}
where, for each grid cell $i$,  $M_i$ is the stellar mass, $R_i$ is the projected radius, and $V_i$ and $\sigma_i$ are the velocity and velocity dispersion. The sum is over the number of grid cells $N$. The calculation of $\lambda_{\rm R}$ is within $2.5\: \rm R_e$.

\begin{figure}
\center
\includegraphics[width=\columnwidth]{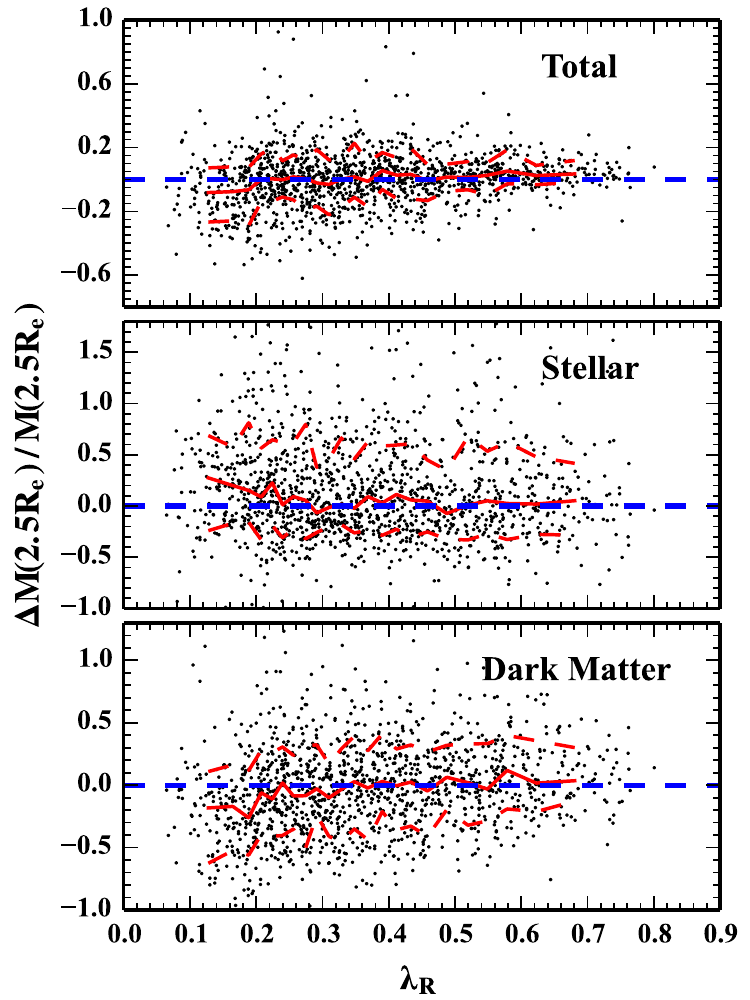}
\caption{ Fractional error of the enclosed total mass (upper), stellar mass (middle), and dark matter (lower)  vs. parameter $\lambda_R$.  The labels and legends are the same as in Fig.~\ref{bias_tri}.
\label{spin}}
\end{figure}

In Fig.~\ref{spin}, we plot the fractional errors in the enclosed stellar, dark matter and total masses vs. the parameter $\lambda_{\rm R}$. In the upper panel of Fig.~\ref{spin}, the scatter of the total mass error significantly decreases as galaxy rotation increases. In the middle and lower panels, there are biases in the recovered stellar and dark matter masses when $\lambda_R$ is below $0.2$. This may be due to the fact that the assumptions for velocity anisotropy are not strong enough for slow rotators. 

\subsubsection{Recovery of the $f_{\rm DM}-M^*$ relationship}
The dark matter fraction ($f_{\rm DM}$) within a given radius is already non-negligible ($\sim 15\%$) at the effective radius \citep{b26, b27}.
A statistical study of this relationship is important for understanding  galaxy evolution.

\begin{figure*}
\center
\includegraphics[width=0.8\textwidth]{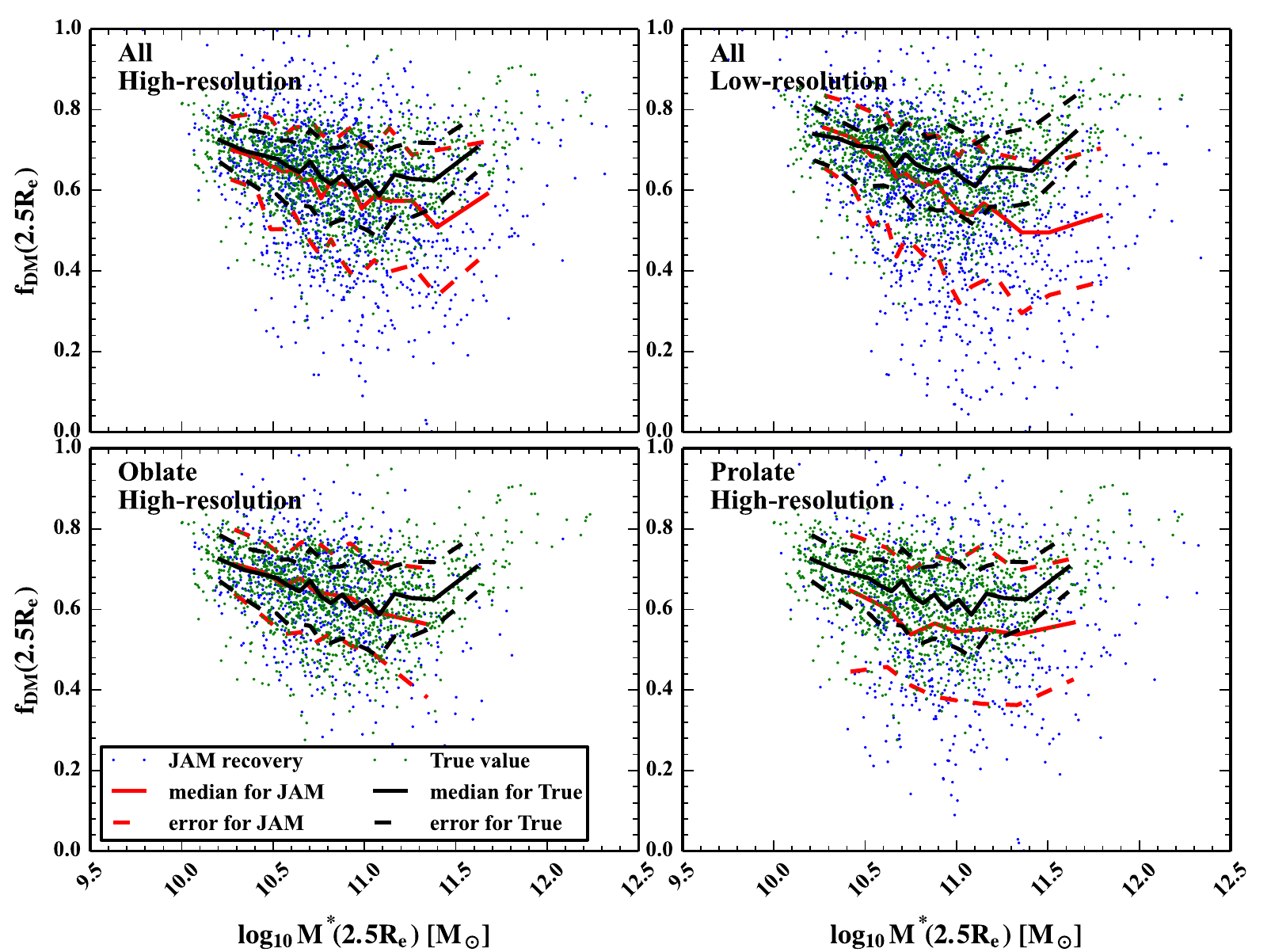}
\caption{Comparison of the $f_{\rm DM}(2.5\: \rm R_e)-M^*(2.5\: \rm R_e)$ relationship between for JAM recovered and true values, where $f_{\rm DM}(2.5\: \rm R_e)=M^{\rm DM}(2.5\: \rm R_e)/M^{\rm T}(2.5\: \rm R_e)$. In each subplot, the green dots are the true values for the whole sample, and the black lines are the median and the $1\sigma$ region in each bin.  The blue dots are the JAM recovered values and the red lines are the median and the $1\sigma$ region in each bin. There are at least 80 points in each bin. Upper left: Recovered relationship for the whole sample with high-resolution images. Upper right: Recovered relationship for the whole sample with low-resolution images. Lower left: Recovered relationship for oblate ($T<0.3$) galaxies with high-resolution images. Lower right: Recovered relationship for non oblate galaxies ($T>0.3$) with high-resolution images.
\label{fr_ml}}
\end{figure*}

In Fig.~\ref{fr_ml}, we compare the recovered $f_{\rm DM}(2.5\: \rm R_e)$ to $M^*(2.5\: \rm R_e)$ relationship from our JAM models with that from the data used.
For the high resolution mock galaxies, we can reproduce the mean $f_{\rm DM}(2.5\: \rm R_e)$ to $M^*(2.5\: \rm R_e)$ relationship for galaxies
less massive than $10^{11}M_{\odot}$. For more massive galaxies, however,  $f_{\rm DM}(2.5\: \rm R_e)$ is underestimated.
We currently believe that this bias is a consequence of  the oblate assumption made in our JAM modelling. In the lower panels, we show the recovered relationship
for oblate galaxies and prolate galaxies separately. While the relationship is recovered for oblate galaxies, prolate galaxies
suffer from an underestimation of the dark matter fraction. 22\% of galaxies with $M^*>10^{11}M_{\odot}$
are prolate or triaxial, and it is these galaxies which cause the bias in the $f_{\rm DM}$ to $M^*$ relationship at the high mass end.

In the same figure, we also show the recovered $f_{\rm DM}$ to $M^*$ relationship from our  low resolution JAM models. Not surprisingly, the bias is even more severe for low-resolution mock galaxies.

\section{Summary of JAM Results}\label{summary}

We have exercised the JAM method using 1413 mock galaxies from the Illustris cosmological hydrodynamic simulation. 
Factors within the method relating to accuracy, degeneracies and biases have been investigated for galaxies with different image resolutions. Notwithstanding that JAM was designed for oblate galaxies, we have included galaxies with different triaxial shapes. We
have also studied whether JAM can recover the $f_{\rm DM}$ to $M^*$ relationship.

For high resolution mock galaxies, we find that JAM reproduces the total mass model within $2.5\: \rm R_e$. The accuracy of the enclosed total mass estimate within $2.5\: \rm R_e$ is $10\%$ for oblate galaxies and $18\%$ for prolate galaxies. However, the JAM method can not disentangle effectively the dark matter and stellar mass fractions for individual galaxies.  From Fig.~\ref{degeneracy_statistical} it can be seen that errors in the stellar mass estimation strongly correlate with errors in dark matter estimation. For the Schwarzschild modelling in \citet{Thomas2007}, the recovery accuracy of the total mass is $3\%$ for oblate galaxies and $20\%$ for prolate galaxies. They also found that while the total mass of the mock galaxies was well recovered, the recovered stellar mass had larger scatters and a significant bias. In \citet{Thomas2007}, all recovered stellar masses are lower than the true values. In our work, both under-estimates and over-estimates exist. It is unclear why this difference has arisen. It may be because their dark matter halo profile has only one free parameter, which limits flexibility to adjust their model.

Though with  large scatters in the stellar and dark matter mass estimates, the recovered stellar masses and dark matter masses are nearly unbiased for oblate galaxies.   
For prolate and triaxial galaxies, stellar masses are on average $18\%$ higher than the true values, while dark matter masses are $22\%$ lower. The bias is mainly a consequence of applying an oblate model to prolate and triaxial galaxies. In reality, it is not straightforward
to determine whether a galaxy is oblate or not (especially for slow rotators), and we need to be mindful of this in constructing galaxy samples.  Due to the galactic mass to shape relationship, this problem may be more severe for high-mass galaxies.

For oblate galaxies with high inclination ($i_{true}>60^{\circ}$), the inclination accuracy is about 5 degrees, with a bias of  2 degrees. The error in the anisotropy parameter $\beta_z$ is 0.11, with a bias of -0.02. However, when the galaxies are near face on, the errors in the  inclination and $\beta_z$ increase and and both are significantly biased.  The accuracy of mass estimates depends on a galaxy's rotation. According to  Fig.~\ref{spin}, the total mass errors are less when the galaxy is a fast rotator.  The recovered stellar and dark matter masses are biased when the galaxy is slow rotator.

The resolution of the image used in the MGE surface brightness fitting also has important effects on the results. When we reduce the resolution from $\rm 0.5kpc/h$ to $\rm 2kpc/h$, the stellar mass bias for all galaxies increases from $5\%$ to $19\%$. For low-resolution images, key information is lost at the centre of galaxies. Accurate and precise estimates of the stellar density profile in the central regions of a galaxy are crucial for reducing the bias in estimating $M^*/L$ and $f_{\rm DM}$.

One important application of JAM could be to recover the $f_{\rm DM}$ to $M^*$ relationship. 
For our high resolution mock galaxies, we recover the mean relation 
below $10^{11}M_{\odot}$ well. As can be seen from Figs.~\ref{bias_tri}, \ref{mass_tri} and \ref{fr_ml}, the relationship for more massive galaxies is not well reproduced due to the fact that galaxies at the high mass end are generally more prolate. For low resolution mock galaxies, 
the relationship is not well reproduced for all masses, again showing the importance of high-resolution images.

\section{Discussion  and Conclusions}\label{discussion}

For this investigation, we have used mock galaxies from the state-of-the-art hydrodynamical Illustris simulation to study the effectiveness and limitations of the JAM modelling technique. Even though our findings are JAM based, we believe they are worthy of consideration for all Jeans equation modelling. Whilst our study explored diverse galaxies from a cosmological simulation, the study has a number of areas that could be improved upon in a future investigation. We discuss these in more detail below.

Although the Illustris simulation can reproduce a variety of observational data, it cannot yet accurately match the stellar mass function (for example), and so it has limitations in terms of resembling real galaxies. The softening length in the simulation for baryons is $710\: \rm pc$, and this may affect the dynamics of stars in the central regions ($\sim2\: \rm kpc$) of galaxies. However most of our galaxies have effective radii larger than $\sim$2 kpc (see Fig.~\ref{stellar_mass}). In addition, we find no correlation between  accuracy of the recovered stellar mass and effective radius, so we believe the effects of the softening length are not significant. This issue can only be conclusively understood when future higher-resolution simulations become available.  Furthermore, better imaging data and more sophisticated modelling techniques such as Schwarzschild and M2M  may better remove this degeneracy. Nevertheless, it will be interesting to examine further the internal dynamical structures of galaxies from simulations, for example the velocity anisotropy parameters as a function of radius, and their variation as a function of stellar mass and galaxy shape.    

In modelling our mock observations, for simplicity, we did not use any data smoothing techniques such as  cloud-in-cell.  We have experimented with cloud-in-cell and find that our results are nearly unchanged for galaxies with median size or larger. Even for the smallest galaxies, where it might be expected that smoothing would have a more significant impact,  only minor differences in stellar mass recovery resulted.

In our work, when setting the $\gamma$ prior boundary, we use additional information from the simulated galaxies (inner star and dark matter slope distributions, see Fig.~\ref{slope_distribution}). Note that even if we set a broader prior for $\gamma$ (e.g. $[-3,0]$), the degeneracy between dark matter and stellar mass does not increase much (scatter increased by $\sim4\%$) compared with the $\sim30\%$ original scatter. The reason we set a narrower prior ($[-1.2,0]$) according to Fig.~\ref{slope_distribution} is to be consistent with \cite{b5}, who used a similar approach to set the prior for their dark matter slope distributions.

In this study, we assumed the whole galaxy has a constant stellar mass-to-light ratio, and this is a commonly adopted approach in dynamical modelling. In fact, hydrodynamical simulations now contain information about star formation and chemical abundances for the stellar populations, thus it will be interesting to incorporate such information in future dynamical modelling, e.g., the $M^*/L$ gradient (Ge et al. 2015, in preparation). Furthermore, it will be interesting to explore and model chemo-dynamical correlations.

We have primarily studied what we can infer if the observational data extends spatially to $2.5\: \rm R_e$. About 1/3 of galaxies from MaNGA will have data to such extent, but the rest will have data only to $1.5\: \rm R_e$. It may be more difficult to infer the $M^*/L$ accurately using such data. Observationally it may be useful to obtain ancillary data at larger radii, for example, using long-slit data or globular clusters \citep{Zhu2014}. 

Our study showed that high-resolution imaging helps to alleviate the stellar mass and dark matter degeneracy for nearby galaxies.  Images captured under good seeing conditions will be very useful for reducing the degeneracy. In this respect, stellar population modelling may also help to (partially) lift the degeneracy. 

In the JAM method, we only used the second order moments of the velocity distribution. In principle, with data from IFU surveys such as MaNGA, higher-order moments (in the form of the Gauss-Hermite coefficients $h_3$ and $h_4$) will be available as well. Orbit-based or particle-based modelling techniques, such as the Schwarzschild and the made-to-measure (M2M) methods, can readily utilise such information to refine mass models. It will be interesting to perform a study where different methods (JAM, M2M and Schwarzschild's method, action-based distribution functions) are compared with each other to identify their strengths, weaknesses and complementarity.
The computational requirements of methods other than JAM may however limit the extent to which such a comparison is achievable.

It is our intention to use JAM next on MaNGA galaxies.  Whether in so doing we will require a prolate version of JAM will become evident at that time.

\section*{Acknowledgements}
We thank Dr. Juntai Shen and Liang Gao for useful discussions and Dr. M. Cappellari for making his MGE and JAM software publicly available. 
We gratefully acknowledge the Illustris team, especially Drs Volker Springel and Lars Hernquist, for early access to their data and for comments on an early draft of the paper.

DDX would like to thank the HITS fellowship.  HYL would like to thank Dr. Junqiang Ge for many useful discussions, and help on dealing with observational data errors.

We performed our computer runs on the Zen high performance
computer cluster of the National Astronomical Observatories,
 Chinese Academy of Sciences (NAOC). This work has
been supported by the Strategic Priority Research Program
``The Emergence of Cosmological Structures" of the Chinese 
Academy of Sciences Grant No. XDB09000000 (RJL
and SM), and by the National Natural Science Foundation of
China (NSFC) under grant numbers 11333003 and 11390372
(SM). RL acknowledges the NSFC (grant No.11303033, 11133003),  and the support from Youth Innovation Promotion Association of CAS. 


\label{lastpage}
\end{document}